\documentclass[prb,amsmath,amssymb,twocolumn,showpacs,floatfix]{revtex4}
\usepackage{graphicx}
\usepackage{dcolumn}
\usepackage{bm}
\usepackage{subfigure}

\begin{document}

\title{Universal inhomogeneous magnetic-field response in the normal state of cuprate high-$T_c$ superconductors}

\author{Z.~Lotfi Mahyari}
\affiliation{Department of Physics, Simon Fraser University, Burnaby, British Columbia, Canada V5A 1S6}
\author{A.~Cannell}
\affiliation{Department of Physics, Simon Fraser University, Burnaby, British Columbia, Canada V5A 1S6}
\author{E.V.L. de Mello}
\altaffiliation[Current Address: ]{Department of Physics, Simon Fraser University, Burnaby, British Columbia, Canada V5A 1S6}
\affiliation{Instituto de F\'{i}sica, Universidade Federal Fluminense, Niter\'{o}i, RJ 24210-340, Brazil}
\author{M.~Ishikado} 
\affiliation{Research Center for Neutron Science and Technology, Tokai, Naka, Ibaraki, Japan 319-1106}
\author{H.~Eisaki}
\affiliation{National Institute of Advanced Industrial Science and Technology, Tsukuba, Ibaraki, Japan 305-8568}
\author{Ruixing Liang}
\affiliation{Department of Physics and Astronomy, University of British Columbia, Vancouver, British Columbia, Canada V6T 1Z1}
\affiliation{Canadian Institute for Advanced Research, Toronto, Canada M5G 1Z8}
\author{D.A.~Bonn}
\affiliation{Department of Physics and Astronomy, University of British Columbia, Vancouver, British Columbia, Canada V6T 1Z1}
\affiliation{Canadian Institute for Advanced Research, Toronto, Canada M5G 1Z8}
\author{J.E.~Sonier}
\affiliation{Department of Physics, Simon Fraser University, Burnaby, British Columbia, Canada V5A 1S6}
\affiliation{Canadian Institute for Advanced Research, Toronto, Canada M5G 1Z8}

\date{\today}

\begin{abstract}
We report the results of a muon spin rotation ($\mu$SR) study of the bulk of Bi$_{2+x}$Sr$_{2-x}$CaCu$_2$O$_{8+\delta}$, as 
well as pure and Ca-doped YBa$_2$Cu$_3$O$_y$, which together with prior measurements reveal a universal inhomogeneous magnetic-field 
response of hole-doped cuprates extending to temperatures far above the critical temperature ($T_c$). 
The primary features of our data are incompatible with the spatially inhomogeneous response being dominated by 
known charge density wave (CDW) and spin density wave (SDW) orders. Instead the normal-state inhomogeneous line broadening is found 
to scale with the maximum value $T_c^{\rm max}$ for each cuprate family, indicating it is 
controlled by the same energy scale as $T_c$. Since the degree of chemical disorder varies widely among the cuprates we have measured, 
the observed scaling constitutes evidence for an intrinsic electronic tendency toward inhomogeneity above $T_c$.
\end{abstract}

\maketitle

\section{Introduction}
Experiments probing the normal state of high-$T_c$ cuprate superconductors have provided evidence for the presence of electronic 
nematicity,\cite{Hinkov:08,Daou:10} fluctuating stripes,\cite{Parker:10} CDW fluctuations,\cite{Ghiringhelli:12,Chang:12a} weak magnetic 
order\cite{Fauque:06,Li:08} and superconducting fluctuations (SCFs) or phase-incoherent 
Cooper pairs.\cite{Corson:99,Xu:00,Gomes:07,Wang:05,Li:10,Dubroka:11} At present there is a quest for commonality amongst these findings, and debate over 
the temperature range of SCFs above $T_c$. 

The initial discovery of a vortex-motion contribution to the Nernst signal in the normal state of cuprate 
superconductors,\cite{Xu:00} advocated the occurrence of phase-fluctuating superconductivity up to temperatures several times $T_c$. This finding was 
subsequently supported by high-resolution torque magnetometry experiments, which detected field-enhanced diamagnetism at equally high 
temperatures.\cite{Wang:05,Li:10} More recently, precursor superconductivity persisting up to 180 K has been inferred from the infrared $c$-axis response of 
$R$Ba$_2$Cu$_3$O$_y$ ($R$ = Y, Gd, Eu).\cite{Dubroka:11} Yet other studies have argued that a noticeable contribution 
of SCFs to the Nernst response of cuprates is present only at temperatures 10 to 25~K above $T_c$.\cite{Daou:10,Cyr:09,Rullier:06} 
A narrow region of SCFs above $T_c$ has also been concluded from AC conductivity studies.\cite{Corson:99,Grbic:11,Bilbro:11}

Distinct from these bulk measurements is a scanning tunneling microscopy (STM) study of the surface of Bi$_{2+x}$Sr$_{2-x}$CaCu$_2$O$_{8+\delta}$ (BSCCO), 
which shows the nucleation of pairing gaps in nanoscale regions persisting to temperatures above the detection of diamagnetism.\cite{Gomes:07} 
Although the lack of consensus about the temperature extent of SCFs above $T_c$ may be attributed to 
varying degrees of sensitivity of different techniques to inhomogeneous superconducting correlations, BSCCO, 
and more so its surface, are highly disordered. Whether the same nanoscale 
electronic inhomogeneity observed at the surface of BSCCO above $T_c$ is also present in the bulk, and whether it has any relevance to other cuprate 
superconductors are questions of fundamental importance.

Expanding on earlier measurements,\cite{Sonier:08,MacDougall:10,Kaiser:12}
we have used transverse-field (TF) $\mu$SR to investigate the situation in the bulk of BSCCO and 
optimally-doped and Ca-doped YBa$_2$Cu$_3$O$_y$ (YBCO) single crystals. The implanted positive muon is a pure local magnetic probe, and relaxation of
the time-dependent TF-$\mu$SR signal results from a distribution of internal magnetic field $n(B)$. The line width of 
$n(B)$ and the corresponding relaxation rate are reduced as the internal magnetic field becomes more uniform, but are also diminished by 
fluctuations of the local field.

\begin{figure}
\centering
\includegraphics[width=8.0cm]{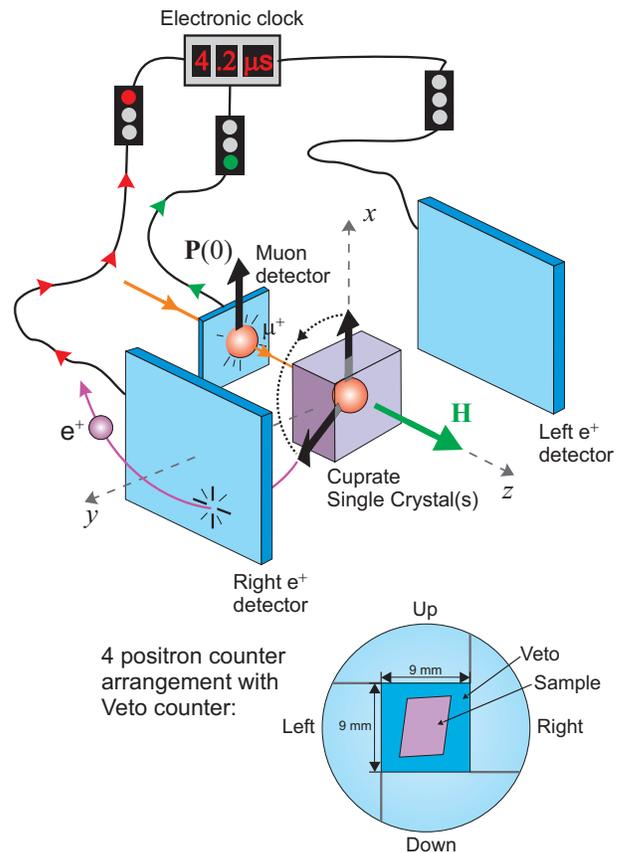}
\caption{(Color online) Schematic of the TF-$\mu$SR experimental arrangement. The $\mu^+$ beam passes through a thin plastic scintillator muon detector with its initial 
muon-spin polarization {\bf P}(0) transverse to the direction of the applied magnetic field {\bf H}, which creates a "start" pulse for an electronic clock. The 
muons are subsequently implanted one-by-one in the sample, where they come to rest and Larmor precess in the local magnetic field {\bf B}. The time evolution of 
the muon spin polarization $P(t)$, which is affected by both static and fluctuating internal magnetic fields, is monitored via the detection of the decay 
positrons (e$^+$). The detection of a decay positron stops the electronic clock, and the corresponding elapsed-time bin of the positron-detector histogram is 
incremented. The positron detectors are arranged in pairs on opposite sides of the sample. In our experiments, two pairs of positron detectors surrounding the 
sample (Left and Right, and Up and Down) were used. In addition, the sample was mounted directly on a "Veto" detector, used to eliminate muons that missed the 
sample from contributing to the TF-$\mu$SR signal. The sample covered approximately 40 to 60~\% of the Veto detector (The lower figure is a depiction of the 
counter arrangement facing the $\mu^+$ beam).
With the exception of the incoming muon detector, all detectors were contained with the sample inside a 
helium-gas flow cryostat. The TF-$\mu$SR asymmetry spectrum $A(t) \! = \! a_0P(t)$ is formed by combining the accumulated histograms of opposing positron detectors.}
\label{fig1}
\end{figure}   

\section{Experimental Details}
The samples are all plate-like single crystals. The pure and Ca-doped YBCO single crystals were grown by a self-flux method in fabricated BaZrO$_3$ crucibles, as described 
elsewhere,\cite{Liang:98} and assembled into mosaics of 6 to 10 single crystals from the same growth batch. Typical sample sizes were 
$5 \! \times \! 5 \! \times \! 0.1-0.2$~mm$^3$. Zero-field (ZF) $\mu$SR measurements on some of the YBCO samples are presented in Ref.~\onlinecite{Sonier:09}. 
The BSCCO samples studied are of similar dimensions, and consist of 1 or 2 single crystals grown by the traveling-solvent-floating-zone method. Sample compositions 
of Bi$_{2+x}$Sr$_{2-x}$Ca$_2$Cu$_2$O$_{8 \! + \! \delta}$ with $x \! = \! 0.1$ to 0.15 were fabricated from powders of Bi$_2$O$_3$, SrCO$_3$, CaCO$_3$, and CuO 
as starting materials. After pre-melting the polycrystalline rod, crystal growth was carried out in air and at a feed speed of 0.15 to 0.20 mm/h for about 3 weeks. 
The doping level was adjusted by tuning the excess oxygen content. The underdoped BSCCO sample was annealed at 570~$^\circ$C under flowing N$_2$ gas with less than 10 ppm 
oxygen concentration for 72 hrs. Overdoping was achieved by annealing at 400~$^\circ$C under an oxygen partial pressure of 2.3 atm for 72 to 250 hrs. The 
optimally-doped BSCCO sample was annnealed in air at 720~$^\circ$C for 24 hrs. A superconducting quantum interference device (SQUID) magnetometer was used for 
measurements of $T_c$. 

\begin{figure}
\centering
\includegraphics[width=9.0cm]{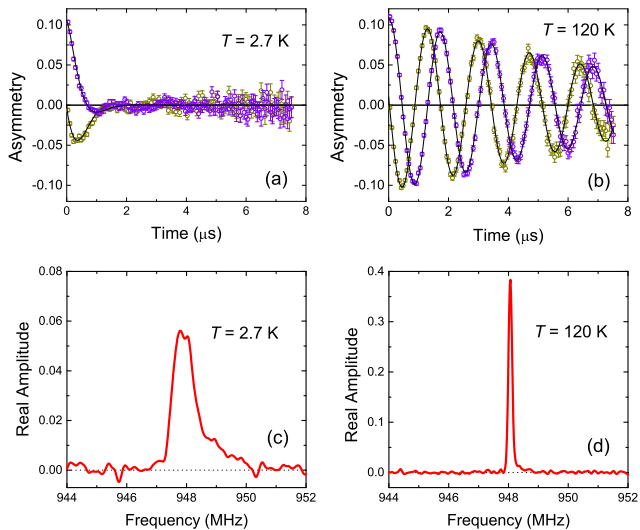}
\caption{(Color online) Representative TF-$\mu$SR signals. (a) TF-$\mu$SR asymmetry spectrum for optimally-doped ($p \! = \! 0.165$) YBCO at $H \! = \! 7$~T and $T \! = \! 2.7$~K. 
The two signals, which differ in phase by $\sim \! 90^\circ$, come from the two pairs of opposing positron detectors shown in Fig.~\ref{fig1} ({\it i.e.} Up and Down, and Left and Right). 
The solid black curves through the data points are fits described in the main text. (b) Same as (a), but for $T \! = \! 120$~K. (c), (d) Fourier transforms 
(with Gaussian apodization) of the asymmetry spectra, which provide visual depictions of the internal magnetic field distribution $n(B)$ sensed by the muon ensemble. 
The frequency (horizontal axis) is related to the local internal magnetic field via the relation $f \! = \! (\gamma_\mu/2 \pi)B$.}
\label{fig2}
\end{figure}

\begin{figure*}
\centering
\includegraphics[width=18.0cm]{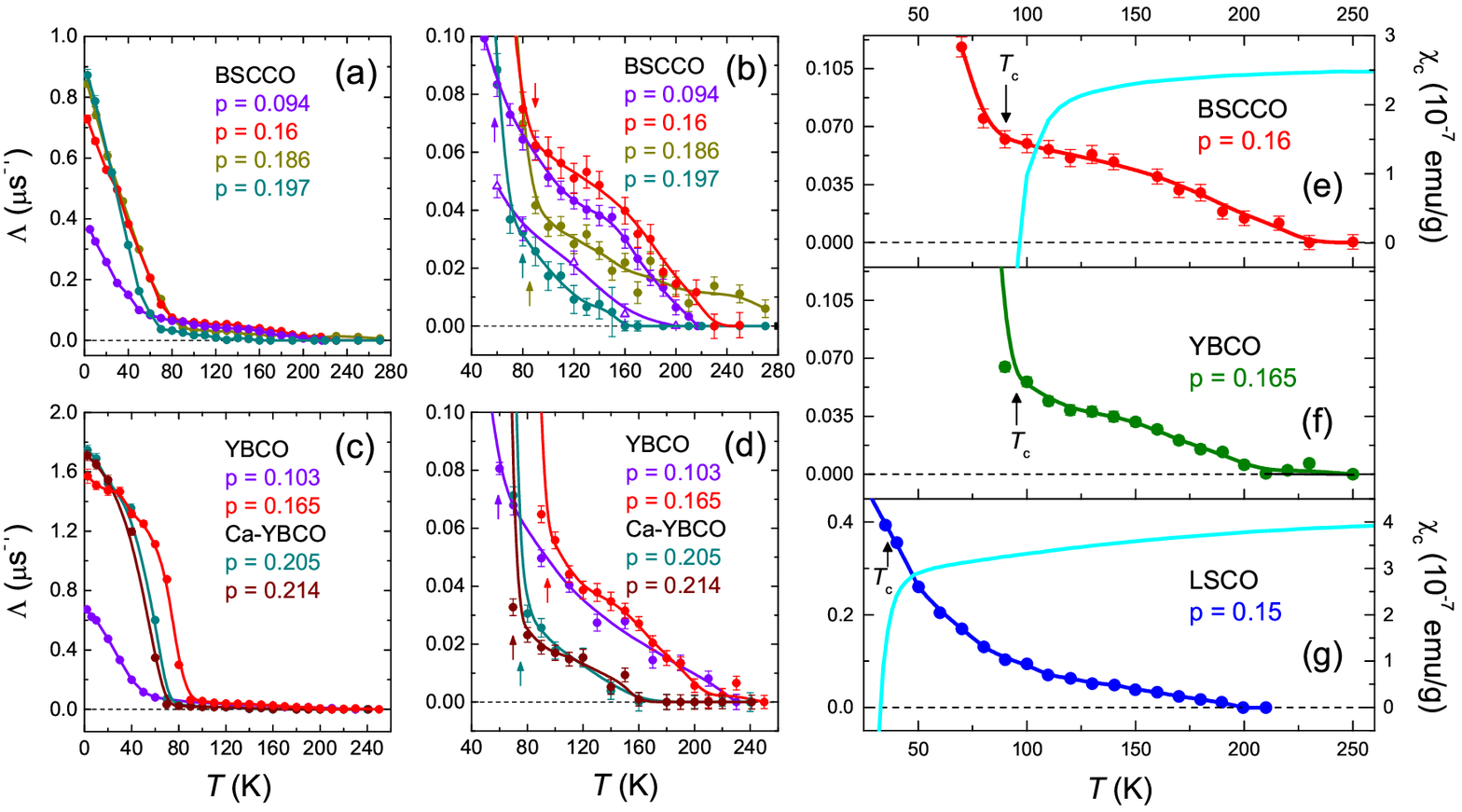}
\caption{(Color online) (a) Temperature dependence of $\Lambda$ at $H \! = \! 7$~T (solid circles) 
for BSCCO. (b) Blow-up of the high-temperature data from (a). Also shown is data for the 
$p \! = \! 0.094$ sample at $H \! = \! 0.5$~T (open purple triangles). The arrows indicate the zero-field 
values of $T_c$ determined by bulk magnetic susceptibility. (c), (d) Representative results for pure and Ca-doped YBCO for $H \! = \! 7$~T.
The data at $p \! = \! 0.103$ is from Ref.~\onlinecite{Sonier:08}. (e), (f), (g) High-temperature 
behavior of $\Lambda$ (solid circles) for optimally-doped BSCCO, YBCO, and LSCO. Also shown 
is the bulk magnetic susceptibility (light-blue curves) for the BSCCO and LSCO samples for $H \! = \! 7$~T
applied parallel to the $c$ axis.}
\label{fig3}
\end{figure*}   

The TF-$\mu$SR experiments were carried out on the M15 surface muon channel at TRIUMF (located in Vancouver, Canada) using the so-called HiTime spectrometer, which 
features ultra-low background and high magnetic field capabilities. The magnetic field {\bf H} was applied parallel to the $c$ axis of the sample by a 7.0~T 
superconducting split-coil solenoid. Nearly 100~\% spin-polarized positive muons were implanted into the sample with the initial muon-spin polarization {\bf P}(0) 
transverse to the direction of the applied field (see Fig.~\ref{fig1}). The muon magnetic moment undergoes Larmor precession at a frequency proportional to the 
local internal magnetic field ({\it i.e.} $\omega \! = \! \gamma_\mu B$, where $\gamma_\mu \! = \! 851.6$~MHz/T is the muon gyromagnetic ratio), which is the vector 
sum of the applied field penetrating the sample, and the dipolar fields associated with the magnetic moments of the host nuclei and electrons. The time evolution of
the muon-spin polarization $P(t)$ reflects the distribution of internal magnetic fields experienced by the muon ensemble, and is monitored through the detection of 
the decay positrons of the implanted muons.

Calibration measurements on pure Ag for $H \! = \! 7$~T and $H \! = \! 0.5$~T show a temperature-independent TF-$\mu$SR signal. Measurements on the 
cuprate samples at temperatures below $T_c$ were performed under field-cooled conditions, to generate the most uniform vortex lattice. At temperatures 
above $T_c$, the TF-$\mu$SR signal for all samples was found to be independent of whether the measurements were recorded under field-cooled or 
zero-field cooled conditions. Typical TF-$\mu$SR signals for YBa$_2$Cu$_3$O$_{6.93}$ are presented in Fig.~\ref{fig2}. For $T \! < \! T_c$ the relaxation of 
the TF-$\mu$SR signal is dominated by the spatial field-inhomogeneity created by the vortex lattice, which results in a broad asymmetric internal magnetic 
field distribution $n(B)$. Conversely, the reduced relaxation observed for $T \! > \! T_c$ is associated with a narrow and symmetric $n(B)$.

\section{Data Analysis and Results}
 The TF-$\mu$SR spectra were fit to
\begin{equation}
A(t) = a_0 P(t) = a_0 G(t) \cos(\omega_\mu t + \phi) \, ,
\end{equation}
where $a_0$ is the amplitude, $\phi$ is the phase angle between the axis of the positron detector and the initial muon-spin polarization {\bf P}(0), 
$\omega_{\mu}$ is the Larmor frequency and $G(t)$ is a function that describes the relaxation of the TF-$\mu$SR signal.
In particular, $G(t) \! = \! G_{\rm nuc}(t)G_{\rm other}(t)$, where $G_{\rm nuc}(t)$ is a temperature-independent function due to the distribution of 
random nuclear dipole fields, and $G_{\rm other}(t)$ is a phenomenological function describing the signal relaxation by other internal sources. 
For YBCO and BSCCO, $G_{\rm nuc}(t) \! =  \! \exp(-\Delta^2t^2)$, as is usually the case for a dense system of randomly oriented magnetic moments. 
However, recent ZF-$\mu$SR measurements show that the nuclear contribution for LSCO deviates somewhat from a pure Gaussian function,\cite{Huang:12} 
and the relaxation of the TF-$\mu$SR signal at $T \! = \! 200$~K has the functional form 
$G(t) \! = \! \exp[-(\Lambda t)^\beta]$, with $1.72 \! \leq \! \beta \! \leq \! 1.87$.\cite{Kaiser:12} Consequently, we assume $G_{\rm nuc}(t) \! = \! \exp(-\lambda t)^n$
for LSCO, where $\lambda$ and $n$ are the values of $\Lambda$ and $\beta$ determined at $T \! = \! 200$~K in Ref.~\onlinecite{Kaiser:12}. 

\begin{figure}
\centering
\includegraphics[width=9.5cm]{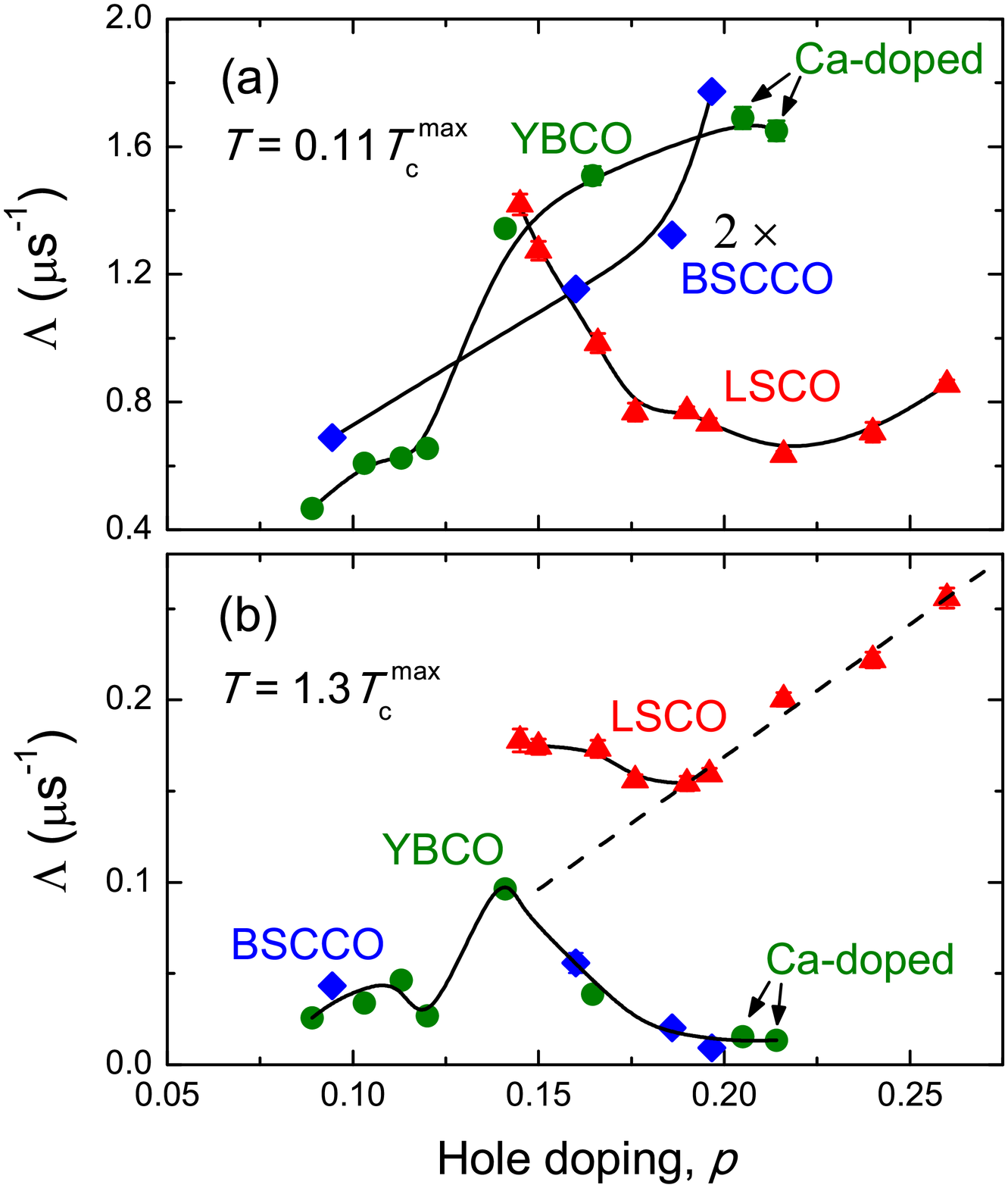}
\caption{(Color online) (a) Doping dependence of $\Lambda$ for $H  \! = \! 7$~T and $T \! = \! 0.11T_c^{\rm max}$, 
where $T_c^{\rm max} \! = \! 90$, 94.1 and 38 K, for BSCCO, YBCO, and LSCO, respectively.  The YBCO 
data for $p \! \leq \! 0.141$ are from Ref.~\onlinecite{Sonier:08}. For visual purposes the relaxation rate for BSCCO 
has been multiplied by a factor of 2. (b) Results for $T \! = \! 1.3T_c^{\rm max}$, where $\Lambda$ is 
a pure exponential relaxation rate. The BSCCO (blue diamonds) and YBCO (green circles) data are for $T \! = \! 120$~K, 
and the LSCO data (solid red triangles) for $T \! = \! 50$~K. Unlike in (a), the BSCCO data is not rescaled. 
The dashed line is a linear fit of the LSCO data for $p \! \geq \! 0.19$, which describes the Curie-like 
contribution.}
\label{fig4}
\end{figure}

For temperatures $T \! < \! 0.5$~$T_c$, the internal magnetic field distribution $n(B)$ for all samples is asymmetric due to an arrangement of vortices, 
and satisfactory fits of the TF-$\mu$SR signal were achieved with $G_{\rm other}(t) \! = \! \exp[-(\Lambda t)^\beta]$, where 
$1.1 \! \leq \! \beta \! \leq 1.8$. However, at higher temperatures $n(B)$ is symmetric, and $G_{\rm other}(t)$ is a pure exponential relaxation function, 
{\it i.e.} $G_{\rm other}(t) \! = \! \exp(-\Lambda t)$. Hence we fit the TF-$\mu$SR signal for all samples above $T_c$ to
\begin{equation}
A(t) = a_0 G_{\rm nuc}(t) \exp(-\Lambda t) \cos(\omega_\mu t + \phi) \, .
\label{asymmetry}
\end{equation}

Figure~\ref{fig3} shows representative data for the temperature dependence of $\Lambda$. Below $T_c$ the dominant contribution to $\Lambda$ is the spatially 
inhomogeneous field created by vortices, which depends on the inverse square of the in-plane magnetic penetration depth $\lambda_{ab}$, 
the anisotropy, and the spatial arrangement of vortices. As shown in Fig.~\ref{fig4}(a), the hole-doping dependence of $\Lambda$ for 
YBCO and BSCCO at low temperatures resembles the doping dependence
of $\lambda_{ab}^{-2}$.\cite{Sonier:07,Anukool:09} The value of $\Lambda$ is significantly smaller for BSCCO, partly because of 
extreme anisotropy that permits significant wandering of the vortex lines along their length.\cite{Brandt:91}

\begin{figure}
\centering
\includegraphics[width=8.0cm]{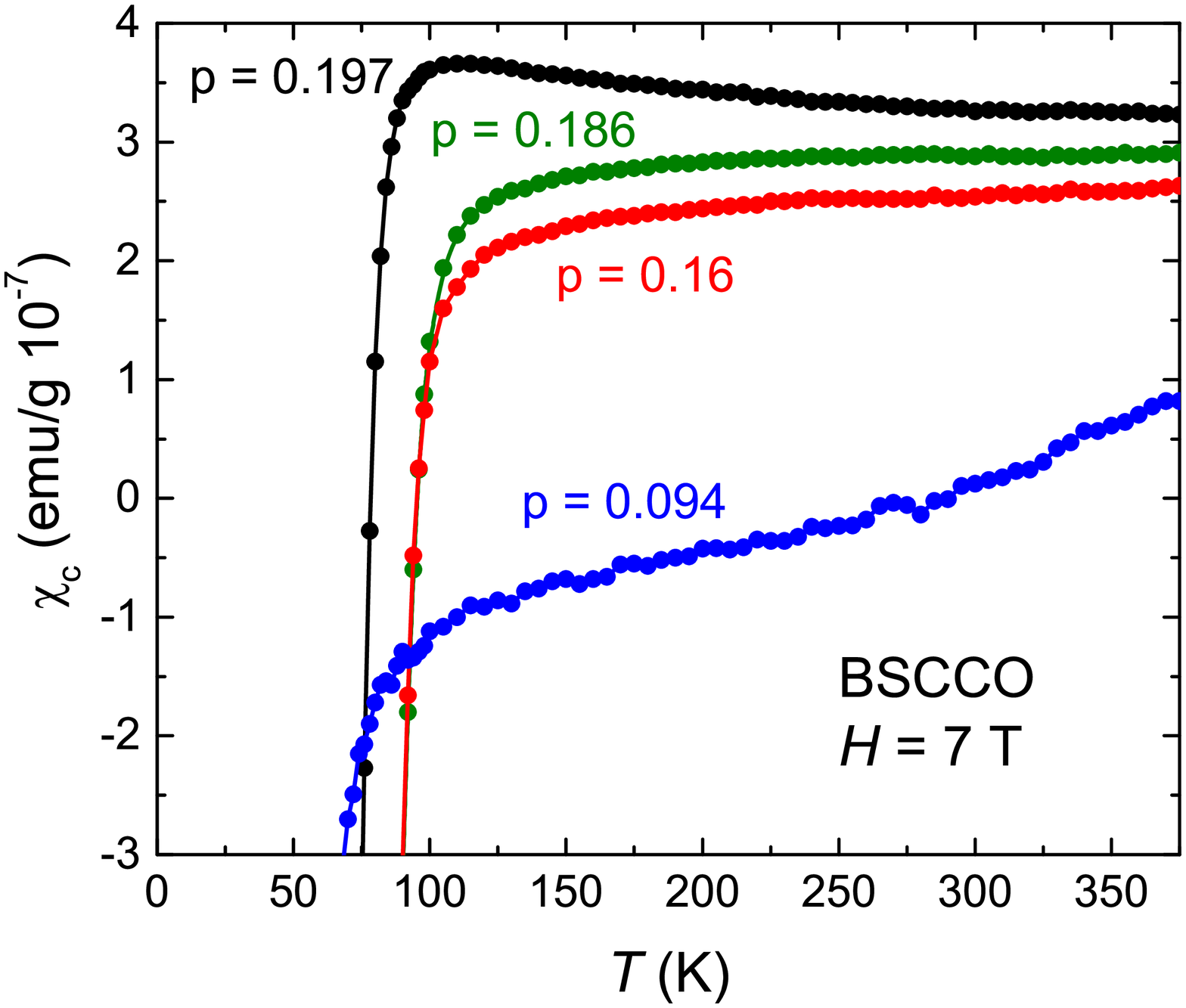}
\caption{(Color online) Temperature dependence of the bulk DC magnetic susceptibility of the $p \! = \! 0.094$, 0.16, 0.186, and 0.197 BSCCO single crystals 
for an external magnetic field $H \! = \! 7$~T applied parallel to the $c$-axis.}
\label{fig5}
\end{figure}
   
For LSCO the observed decrease of $\Lambda$ with increased hole-doping in the range 
$0.145 \! \leq \! p \! \leq \! 0.176$ [Fig.~\ref{fig4}(a)] is opposite to the behavior 
of $\lambda_{ab}^{-2}(p)$.\cite{Lemberger:11} This is explained by
a recent small-angle neutron scattering study of LSCO showing enhanced vortex-lattice disorder below 
$p \! \sim \! 0.18$, near a state of coexisting superconductivity and SDW order.\cite{Chang:12b}
Such random frozen disorder of the rigid vortex lines in LSCO broadens $n(B)$ and enhances $\Lambda$. 
The upturn of $\Lambda$ at higher doping also opposes the behavior of $\lambda_{ab}^{-2}(p)$, which decreases
beyond $p \! \sim \! 0.21$.\cite{Lemberger:11} This is due to Curie-like paramagnetism 
that dominates $\Lambda$ in heavily-overdoped LSCO, but is also present in the underdoped regime.\cite{Kaiser:12,MacDougall:10} 
This contribution is evident in the temperature dependence of $\Lambda$ for optimally-doped LSCO 
presented in Fig.~\ref{fig3}(g). There is a similar Curie term discernible in the bulk magnetic susceptibility
of the $p \! = \! 0.197$ BSCCO sample (see Fig.~\ref{fig5}), which may contribute somewhat to $\Lambda(T)$ at this doping.
The bulk magnetic susceptibility of our BSCCO samples at $H \! = \! 7$T resembles previous high-field measurements.\cite{Watanabe:00}
This includes the Curie-like contribution to the $p \! = \! 0.197$ sample, which is a common feature of heavily-overdoped cuprates.

\begin{figure}
\centering
\includegraphics[width=8.5cm]{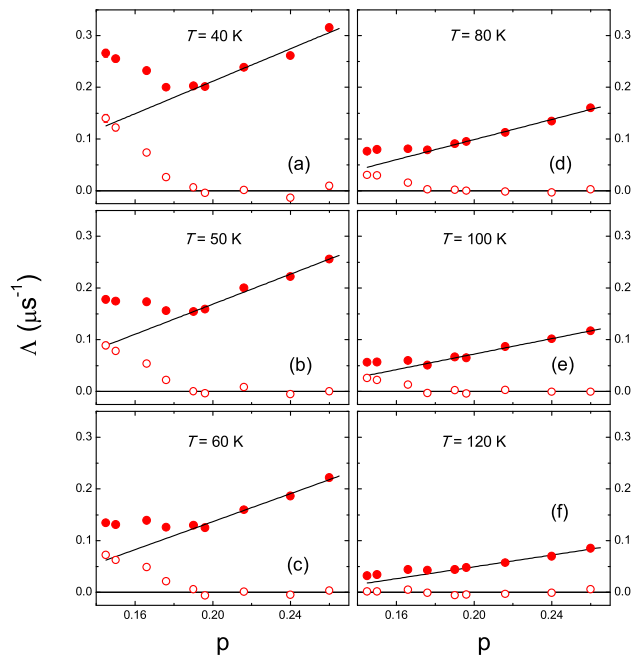}
\caption{(Color online) Hole-doping dependence of the exponential relaxation rate $\Lambda$ for LSCO from fits to Eq.~(\ref{asymmetry}) (solid red circles). 
The solid black line in each panel is the best fit of a straight line to the data at $p \! \geq \! 0.19$. Subtracting the straight line fit from $\Lambda$ 
yields the open red circles.}
\label{fig6}
\end{figure}

Shifting attention to the normal state, Fig.~\ref{fig4}(b) shows a comparison of the hole-doping dependence
of $\Lambda$ at $T \! = \! 1.3T_c^{\rm max}$. Despite BSCCO possessing a higher degree of chemical disorder
than YBCO, there is good agreement between the data sets for these two compounds.
As shown previously,\cite{Kaiser:12} the Curie-like contribution to $\Lambda$ for LSCO exhibits a
dominant $p$-linear dependence above $T_c$, with a slope ($d \Lambda/dp$) that weakens with increasing $T$.
Measurements at higher temperature show that the $p$-linear contribution extends to lower doping.
Figure~\ref{fig6} shows the hole-doping dependence of $\Lambda$ for LSCO above $T_c$ (solid red circles) obtained from fits of the TF-$\mu$SR asymmetry spectra of 
Ref.~\onlinecite{Kaiser:12} to Eq.~(\ref{asymmetry}). Here we stress that the values of $\Lambda$ differ from those in Ref.~\onlinecite{Kaiser:12} due to the removal 
of the nuclear dipole contribution to $G(t)$, which must be done to compare the residual relaxation rate to the values for YBCO and BSCCO.
The linear growth of $\Lambda$ above $p \! = \! 0.19$ at $T \! = \! 40$~K is visible below $p \! = \! 0.19$ at high temperatures, and arises from 
the Curie-like paramagnetism.\cite{MacDougall:10,Kaiser:12} Subtracting this linear contribution from the hole-doping dependence of $\Lambda$ at each temperature 
(yielding the open red circles in Fig.~\ref{fig6}) results in a collapse of the $\Lambda$ data for LSCO
onto a universal curve shared with BSCCO and YBCO (Fig.~\ref{fig7}). Note that for the range of comparison
$0.145 \! \leq \! p \! \leq 0.176$, the applied field of 7~T corresponds to approximately one tenth of 
the upper critical field $H_{c2}$ for all three compounds.\cite{Rourke:11,Grissonnanche:13} The universal
behavior for $\Lambda$ suggests that $T_c$ and the source of the inhomogeneous magnetic-field response above $T_c$ are 
controlled by the same energy scale.

\section{Discussion}
It may be surprising to some that a significant inhomogeneous magnetic response occurs in YBCO above $T_c$, given that the mean free path of
optimally-doped YBCO single crystals has been reported to be as large as 4~$\mu$m.\cite{Hosseini:99} However, this estimate is at low 
temperatures far below $T_c$ and in zero applied magnetic field. One of the reasons that the mean free path is so long at low temperatures is 
that the superconducting gap reduces the available phase space for impurity scattering.\cite{Nunner:05} 
In fact there is a dramatic increase of the scattering rate of
optimally-doped YBCO above 20~K,\cite{Hosseini:99} indicating a large reduction of the mean free path with increasing temperature well
before the normal state is reached. In addition, experiments observing quantum oscillations in the electrical resistance of underdoped YBCO 
indicate a significant reduction of the mean free path in the presence of an applied magnetic field.\cite{Doiron:07} For example, at 35~T, the mean free path 
of YBa$_2$Cu$_3$O$_{6.5}$ at low temperatures is roughly estimated to be only 0.016~$\mu$m.\cite{Jaudet:09} 

\begin{figure}
\centering
\includegraphics[width=8.5cm]{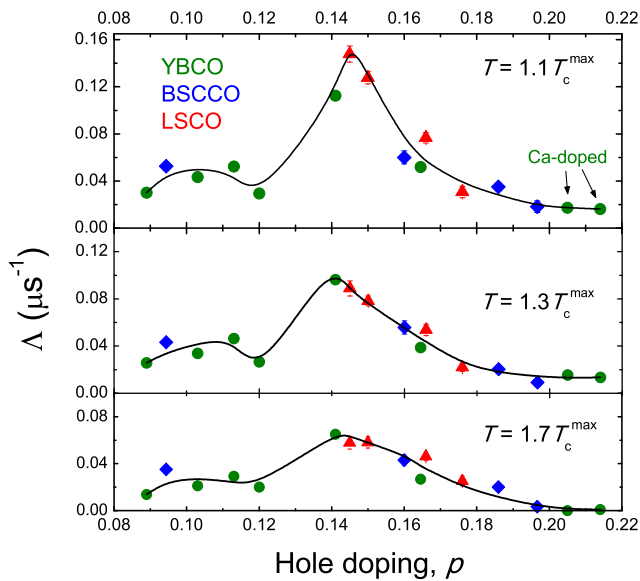}
\caption{(Color online) Doping dependence of $\Lambda$ for $T \! = \! 1.1$, 1.3 and 1.7~$T_c^{\rm max}$. The $p$-linear 
dependent contribution to the data for LSCO has been subtracted, as described in the main text. The solid curves are guides to the eye.}
\label{fig7}
\end{figure}

Previous measurements on YBCO revealed a clear reduction of $\Lambda$ near $p \! = \! 0.12$,\cite{Sonier:08} where 
recent neutron scattering measurements for $H \! = \! 0$ show enhanced incommensurate CDW correlations,\cite{Ghiringhelli:12}
and nuclear magnetic resonance experiments at $H \! = \! 28.5$~T show 
static commensurate CDW order.\cite{Wu:11} Enhanced incommensurate SDW order is also observed in LSCO 
near $p \! = \! 0.12$.\cite{Wakimoto:01} However, slowing down of charge or spin fluctuations should
increase the value of $\Lambda$. Moreover, while $\Lambda$ above $T_c$ increases with $H$ [Fig.~\ref{fig3}(b)],
even fields well in excess of 7~T have no effect on CDW or SDW fluctuations above $T_c$.\cite{Chang:12a,Lake:01}
Hence CDW or SDW correlations, which are confined to a narrow range of doping and seem to compete with 
superconductivity, cannot be the primary source of the residual inhomogeneous line broadening shown in Fig.~\ref{fig7} 
that is observed over the wide doping range $0.089 \! < \! p \! < \! 0.214$.
Additionally, we point out that the muon has no electric quadrupole moment and 
hence does not sense CDW order directly. What it can detect is modifications of the second moment of the distribution 
of nuclear dipolar magnetic fields by quadrupole interactions of the host nuclei with the local electric-field gradient (EFG)
asociated with CDW order. But this is only true in zero or weak applied magnetic field. Although 
the dipolar fields of the nuclei are dependent on both their electric and magnetic interactions,\cite{Hartmann:77} when the 
applied magnetic field is strong (as in our experiments), the external field direction 
becomes a natural quantization axis for the nuclear spin system. In this case the second moment of the nuclear dipolar field
distribution is independent of the magnitude of the external field and is not altered by changes in the local EFG. 

\begin{figure}
\centering
\includegraphics[width=9.0cm]{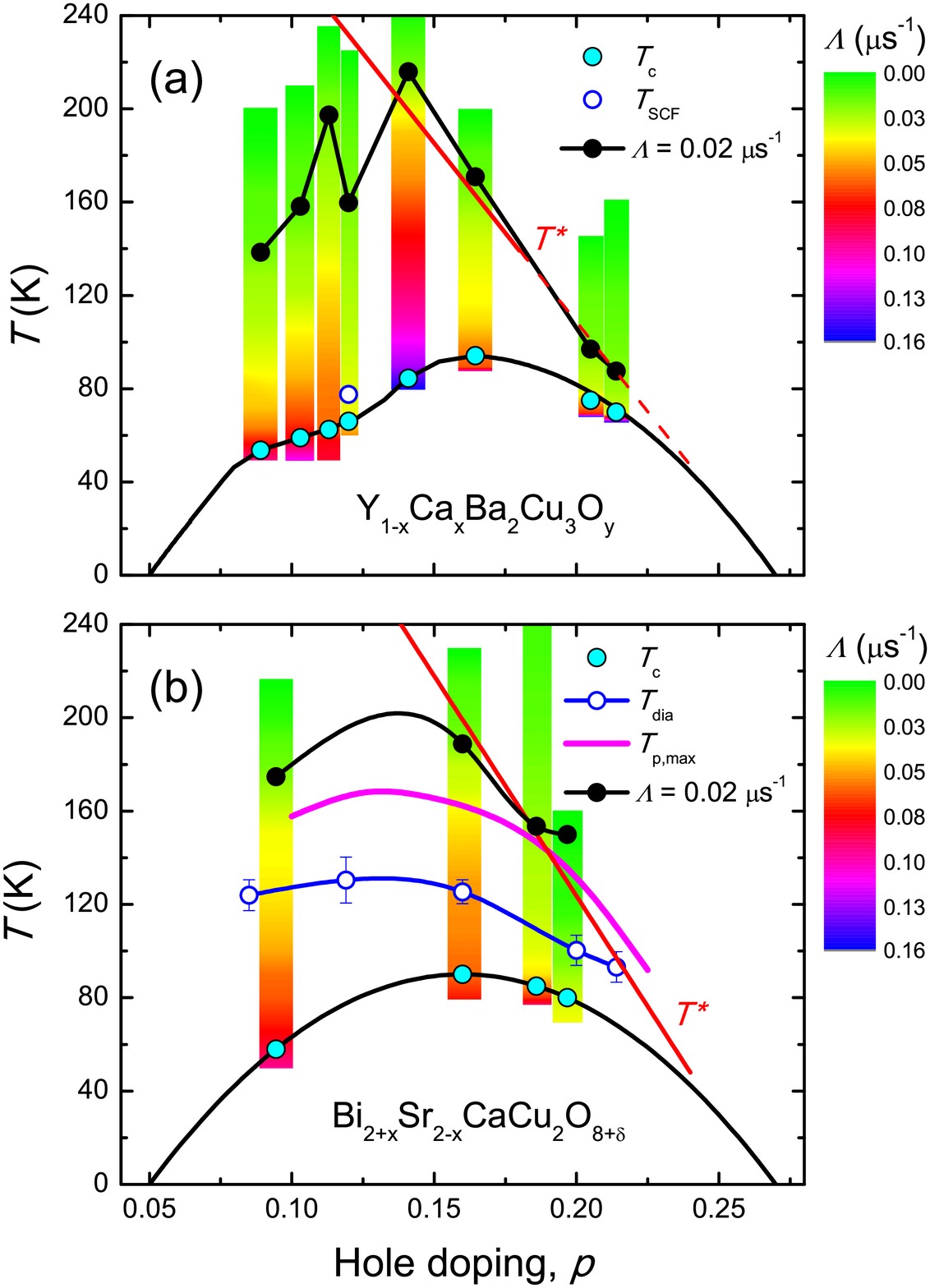}
\caption{(Color online) (a) Contour bar graph of the variation of $\Lambda$ with temperature and hole 
doping in pure and Ca-doped YBCO for $H \! = \! 7$~T, achieved by interpolation of the $\Lambda$ vs. $T$
data sets. The black connected circles indicate the arbitrary value $\Lambda \! = \! 0.02$~$\mu$s$^{-1}$. 
Also shown is the onset temperature $T_{\rm SCF}$ for SCFs at 
$p \! = \! 0.12$ inferred from Nernst-effect measurements,
and $T^*$ (solid red line) extrapolated to higher doping (dashed red line) from Ref.~\onlinecite{Daou:10}. 
(b) Similar schematic phase diagram for BSCCO, where $T_{\rm dia}$ is the temperature at which torque magnetometry 
measurements detect the onset of diamagnetism\cite{Wang:05} and $T^*$ is from Ref.~\onlinecite{Ren:12}.
The $T_{p, max}$ curve indicates the temperature above which STM measurements\cite{Gomes:07}
on BSCCO find less than 10~\% of the sample containing nanometer-sized regions with pairing gaps.}
\label{fig8}
\end{figure}   

Figure~\ref{fig8} shows that $\Lambda$ persists above the pseudogap temperature $T^*$ at high doping,
but vanishes below $T^*$ in the underdoped region, indicating an origin not solely related to 
the pseudogap phase. Instead the doping dependence and universal scaling of $\Lambda$ with $T_c^{\rm max}$
suggest that the inhomogenous field response above $T_c$ is associated with superconducting
correlations. Like the vortex Nernst signal\cite{Xu:00} and diamagnetism observed by torque magnetometry,\cite{Wang:05,Li:10} 
$\Lambda$ for $T \! > \! T_c$ is reduced with increasing temperature and enhanced by the external field. 
The diamagnetism and vortex Nernst signal have been attributed to SCFs, which are
observed over a narrow range of temperature above $T_c$ on a frequency scale of 0.01 - 10~THz.\cite{Corson:99,Grbic:11,Bilbro:11}
At somewhat higher temperatures, our experiments show $\Lambda$ varying on the order of 0.01 - 0.1~$\mu$s$^{-1}$
(Fig.~\ref{fig7}). Considering for a moment $\Lambda$ to be a dynamic relaxation rate due to fast fluctuations 
of the local field, $\Lambda \! = \! \gamma_\mu^2 \langle (\delta B)^2 \rangle/\nu$, where $\nu$ is the fluctuation
frequency and $\langle (\delta B)^2 \rangle$ is the second moment of $n(B)$ in
the static limit $(\nu \! \rightarrow \! 0)$. For 0.01~THz fluctuations, the range $0.01 \! < \! \Lambda \! < \! 0.1$~$\mu$s$^{-1}$
corresponds to a static line width of $0.012 \! < \! \sqrt{\langle (\delta B)^2 \rangle} \! < \! 0.037$~T, which
exceeds the line width of $n(B)$ associated with the frozen vortex lattice below $T_c$. Hence the vortex liquid
inferred from the Nernst signal above $T_c$ is not detectable by $\mu$SR. This conclusion is supported by a $\mu$SR 
study of the sizeable vortex-liquid regime of BSCCO below $T_c$,\cite{Lee:95} where the TF-$\mu$SR 
line width is severely narrowed by thermal vortex fluctuations. In stark contrast to the field dependence of 
$\Lambda$ observed above $T_c$ [Compare the data of $p \! = \! 0.094$ for $H \! = \! 0.5$~T and $H \! = \! 7$~T in
Fig.~\ref{fig3}(b)], the TF-$\mu$SR line width previously measured in the vortex-liquid phase of BSCCO is reduced
by stronger applied magnetic field.

Although homogeneous SCFs do not produce $\mu$SR line broadening, spatially varying SCFs will. The local 
pairing observed by STM\cite{Gomes:07} on BSCCO above $T_c$ is characterized by a distribution
of gap sizes and a partial suppression of the density of states at the Fermi energy $N(E_F)$, which vanish
inhomogeneously with increasing $T$. In a conventional metal the Pauli susceptibility $\chi_0$ 
associated with the conduction electrons is proportional to $N(E_F)$. Spatial variations in the magnitude 
of $\chi_0$ associated with nanoscale regions of pairing cause inhomogeneous broadening via the hyperfine 
coupling between the $\mu^+$ and the spin polarization of the surrounding conduction electrons. Yet any 
such contribution to $\Lambda$ must be minor, since above $T_c$ the depletion of $N(E_F)$ in
the underdoped regime is dominated by the spatially inhomogeneous pseudogap. 

Regular inhomogeneous regions of SCFs may occur in the bulk from a competition with some other kind 
of fragile order. One candidate is fluctuating stripes, which are characterized by
a dynamical unidirectional modulation of charge, or both spin and charge.\cite{Kivelson:03} In this environment inhomogeneous
line broadening will result from muons experiencing distinct time-averaged local fields $\langle B(t) \rangle$
in different parts of the sample. Such will be the case for muons stopping inside regions with different 
strengths of fluctuation diamagnetism. However, even if the regions in which SCFs persist are comparable
throughout the sample, inhomogeneous line broadening results from muons stopping in intermediate
or surrounding areas. These muons experience the expelled time-averaged field, which diminishes in magnitude
with increased stopping distance away from the diamagnetic regions. As shown in Fig.~\ref{fig8}(b), $\Lambda$ tracks the
pairing gap coverage in BSCCO observed by STM --- where the latter is presumably similar regardless of
whether the regions of SCFs maintain a regular pattern or are broken up into irregular-shaped patches by disorder. 

While our findings strongly favor an interpretation involving inhomogeneous SCFs, current instrument detection limits
prevent us from determining whether the magnetic response above $T_c$ is diamagnetic. Even so, our results 
show that there is an intrinsic electronic propensity toward inhomogeneity in the normal-state of 
high-$T_c$ cuprate superconductors, where even weak disorder of the kind found in YBCO is sufficient
to spatially pin the non-uniform electron liquid.

We gratefully acknowledge informative and insightful discussions with S.A. Kivelson, W.N. Hardy, A. Yazdani and L. Taillefer.
This work was supported by the Natural Sciences and Engineering Research Council of Canada, the Canadian 
Foundation for Innovation, the Canadian Institute for Advanced Research, and CNPq/Brazil.

\end{document}